\documentclass{PoS}

\usepackage{amsmath}
\usepackage{gensymb}
\usepackage{placeins}
\usepackage{url}

\title{Multi-wavelength observation of MAXI~J1820+070 with MAGIC, VERITAS and H.E.S.S.}

\ShortTitle{MWL observation of MAXI~J1820+070}

\author{\speaker{J.~Hoang}$^{1,\dag}$, E.~Molina$^{2,\dag}$, M.~L\'{o}pez$^{1,\dag}$, M.~Rib\'{o}$^{2,\dag}$, O.~Blanch$^{3, \dag}$, J.~Cortina$^{4,\dag}$, G.~Maier$^{5,\ddagger}$, N.~Park$^{6,\ddagger}$, M.~de~Naurois$^{7,\S}$, E.~de~Ona~Wilhelmi$^{5,8,\S}$, J.P.~Ernenwein$^{9,\S}$, D.~Malyshev$^{10,\S}$, A.~Mitchell$^{11,\S}$, S.~Ohm$^{5,\S}$, R.~Zanin$^{12,\S}$,\, on behalf of the MAGIC\thanks{https://magic.mpp.mpg.de},\,  VERITAS\thanks{https://veritas.sao.arizona.edu},\, and H.E.S.S.\thanks{https://www.mpi-hd.mpg.de/hfm/HESS \newline
\hspace*{7mm} For collaborations lists see PoS(ICRC2019)1177}\,  Collaborations.\\
       {\footnotesize $^{1}$Instituto de Part\'{i}culas y Cosmolog\'{i}a (IPARCOS), Universidad Complutense de Madrid, Madrid, Spain\\ 
       $^{2}$Institut de Ci\`encies del Cosmos (ICCUB), Universitat de Barcelona (IEEC-UB), Barcelona, Spain\\ 
       $^{3}$Institut de F\'{i}sica d'Altes Energies (IFAE), Barcelona, Spain\\ 
       $^{4}$Centro de Investigaciones Energ\'{e}ticas, Medioambientales y Tecnol\'{o}gicas (CIEMAT), Madrid, Spain\\ 
       $^{5}$Deutsches Elektronen-Synchrotron (DESY), Zeuthen, Germany\\
       $^{6}$Wisconsin IceCube Particle Astrophysics Center, Wisconsin, USA\\ 
       $^{7}$Laboratoire Leprince-Ringuet, \'{E}cole Polytechnique, CNRS/IN2P3, Palaiseau, France\\
       $^{8}$Institute of Space Sciences (CSIC-IEEC), Barcelona, Spain\\
       $^{9}$Aix Marseille Universit\'{e}, CNRS/IN2P3, CPPM, Marseille, France\\ 
       $^{10}$Institut f{\"u}r Astronomie \& Astrophysik, Universit{\"a}t T{\"u}bingen, T{\"u}bingen, Germany\\
       $^{11}$Physik-Institut, Universit\"{a}t Z\"{u}rich, Z\"{u}rich, Switzerland\\
       $^{12}$Max-Planck-Institut f{\"u}r Kernphysik, Heidelberg, Germany\\
       
       E-mail: \email{kimhoang@ucm.es}}}

\abstract{%MAXI~J1820+070 is a new Low Mass X-ray Binary (LMXB) discovered by the Monitor of All-sky X-ray Image (MAXI) on 2018 March 11. It is the counterpart of ASASSN-18ey, discovered on 2018 March 6 by the All-Sky Automated Survey for SuperNovae (ASAS-SN). This source completed its state transition in 2018, having traced the typical ''q-shaped'' curve on the Hardness-Intensity Diagram (HID). Multi-wavelength (MWL) observation of MAXI~J1820+070 provides valuable insights into the mechanisms producing emission across the electromagnetic spectrum from this class of transients.
MAXI~J1820+070 is a new low-mass microquasar hosting a black hole recently discovered in X-rays by the MAXI instrument. It is the counterpart of ASASSN-18ey, discovered in optical a few days before by ASAS-SN. This source underwent a major outburst in 2018, during which it completed the typical "q-shaped" path in the hardness-intensity diagram. MAGIC, VERITAS and H.E.S.S. gamma-ray telescopes observed the sky position of MAXI~J1820+070 for a total of more than 90 hours in 2018. In addition, some observations were carried out using MAGIC Central Pixel - a dedicated central pixel capable of detecting fast optical signals (10 kHz sampling rate, peak sensitivity in the U-band). This contribution presents the methods used to search for transient optical and very high energy gamma-ray emission from MAXI~J1820+070, as well as the latest results in these energy ranges.}
\FullConference{36th International Cosmic Ray Conference (ICRC2019)\\
		        July 24th - August 1st, 2019\\
	        	Madison, WI, U.S.A.}

\begin{document}
\addtocounter{footnote}{-3} %% we remove 3 as the speaker, copyright and sissa site are 3 footnotes.
%% doing this, footnotes in the document start at 1
%#######################################
\section{Introduction}
\vspace*{-0.1cm}
Low-mass microquasars (LMMQs) are X-ray binary systems consisting of a compact object (CO) and a companion star with a mass below $\sim 10 M_\odot$. The former can be either a black hole or a neutron star, and accretes material from the companion, which has exceeded its Roche lobe, through an accretion disk. The combined action of the accretion process, the CO rotation and the magnetic field leads to the formation of bipolar jets launched from the CO. These jets are powerful sites for acceleration of high-energy particles, which in turn emit broadband non-thermal radiation from radio to X- or gamma-rays. The reader is referred to \cite{mirabel99} for a classical review on microquasars. 

Some LMMQs are transient sources, which go over periodic outbursts that increase their luminosity by several orders of magnitude and which may last from a few weeks to several months. Those LMMQs hosting a black hole (BH) can be mainly classified into two different categories according to their X-ray spectra: the hard state (HS) and the soft state (SS) (e.g. \cite{fender16,remillard06}). In the HS, most of the X-ray luminosity has a non-thermal origin identified by a dominant power-law component in the spectrum. This non-thermal emission likely originates from inverse Compton (IC) scattering of low energy photons in the accretion disk by electrons in a hot corona surrounding the BH. Nevertheless, a scenario in which the synchrotron emission from the jets contributes significantly to the X-ray luminosity cannot be ruled out~\cite{fender16}. The SS has an associated X-ray emission dominated by the thermal radiation from the hot inner regions of the accretion disk. In this state, the jets are much weaker or non-existent at all. In a typical outburst for a LMMQ with a BH, the source normally completes the HS--SS--HS cycle, going through short-lived intermediate states (IM) during the transitions.

MAXI~J1820+070 is a microquasar discovered in X-rays by the Monitor of All-sky X-ray Image (MAXI) \cite{matsuoka09} in March 2018, a few days after its optical detection by the All-Sky Automated Survey for Supernovae (ASAS-SN) \cite{tucker19}. Soon after its discovery, the source reached a flux in the $15-50$~keV energy range of $\sim 4$ times that of the Crab nebula in the same energy band, measured by the Swift's Burst Alert Telescope (BAT), and underwent a HS--SS transition in July, reported by the Neutron Star Interior Composition Explorer (NICER)~\cite{2016SPIE.9905E..1HG}. The exceptionally high X-ray flux led to follow-up observations and multiwavelength campaigns joined by many instruments in a broad wavelength range, from radio to gamma-rays. The distance to MAXI~J1820+070 was set to $\sim 3.5$~kpc from \textit{Gaia} data \cite{gandhi19,bailer18}. The mass of the compact object and the companion star are estimated to be $\sim 5 M_\odot$ \cite{Torres} and $\lesssim 1 M_\odot$ \cite{tucker19}, respectively,  classifying MAXI~J1820+070 as a LMMQ.

Since its discovery, MAXI~J1820+070 stayed in the HS for almost 4 months, and subsequently exhibited a rapid X-ray spectral softening in July 2018, entering the SS. In September 2018, X-ray spectral hardening was observed, and the source eventually returned to the HS shortly before going into quiescence and setting an end to the outburst. The evolution of the hardness ratio (HR) of hard to soft X-ray photons can be observed in the top panel of Figure~\ref{fig:mwl}, where the fast decrease and increase of the HR correspond to the HS--SS and the SS--HS transitions, respectively. A hardness-intensity diagram (HID) is also shown in Fig.~\ref{fig:hid}

In this work, we report the observational results obtained in the Very High Energy (VHE) gamma-ray range by the three current-generation Cherenkov telescope arrays: the High Energy Stereoscopic System (H.E.S.S.), the Major Atmospheric Gamma Imaging Cherenkov Telescopes (MAGIC), and the Very Energetic Radiation Imaging Telescope Array System (VERITAS). Optical data from the central pixel of the MAGIC-II telescope are also shown. Observations are described in Section~\ref{observations}, data analysis and results are detailed in Sect.~\ref{results}, and a discussion is given in Sect.~\ref{discussion}.

%#######################################
\section{VHE Instruments and Observations}\label{observations}
\vspace*{-0.2cm}
Observations of MAXI~J1820+070 were carried out by H.E.S.S., MAGIC and VERITAS. H.E.S.S. is an array of five telescopes located in the Khomas Highland, Namibia ($23\degree$S, $16\degree$E, 1800~m above the sea level). Four of these telescopes have a 12-m diameter dish and a $5\degree$ field of view (FoV), and the fifth telescope has a 28-m diameter dish and a $3.2\degree$ FoV. MAGIC is a system of two 17-m diameter telescopes, with a $3.5\degree$ FoV, located at the Roque de los Muchachos Observatory in La Palma, Spain ($29\degree$N, $18\degree$W, 2200~m above the sea level). It has a sensitivity of $\sim 0.7\%$ of the Crab nebula flux in 50h above 180~GeV \cite{magic16}. Additionally, the Central Pixel of the MAGIC-II telescope was modified to allow simultaneous observations in both VHE and optical. This instrument is able to detect isolated 1-ms optical flashes as faint as $\sim 13.4$~mag in the blue wavelength \cite{cpix_icrc}. VERITAS consists of four 12-m diameter telescopes with a $3.5\degree$ FoV in Southern Arizona, USA ($32\degree$N, $111\degree$W, 1270~m above the sea level), and is sensitive to  1\% of the Crab Nebula flux within 25 hours~\cite{veritas_performance}. 

\begin{table}[h]
 \begin{center}
	\caption{H.E.S.S., MAGIC and VERITAS VHE Observation conditions.}
	\begin{tabular}{c c c c }
    \hline \hline
	Instrument	& Wobble offset [deg]	& Zenith [deg]  & Median zenith [deg] \\
    \hline
	H.E.S.S.	& 0.7		        & 30 - 60     & 33 \\
    MAGIC       & 0.4 	            & 20 - 60     & 32 \\
    VERITAS     & 0.5               & 20 - 60     & 32 \\
    \hline
    \end{tabular}
    \label{tab:observation_conditions}
 \end{center}
 \vspace*{-0.3cm}
\end{table}

\begin{figure}
	\centering
    \includegraphics[width=0.9\linewidth]{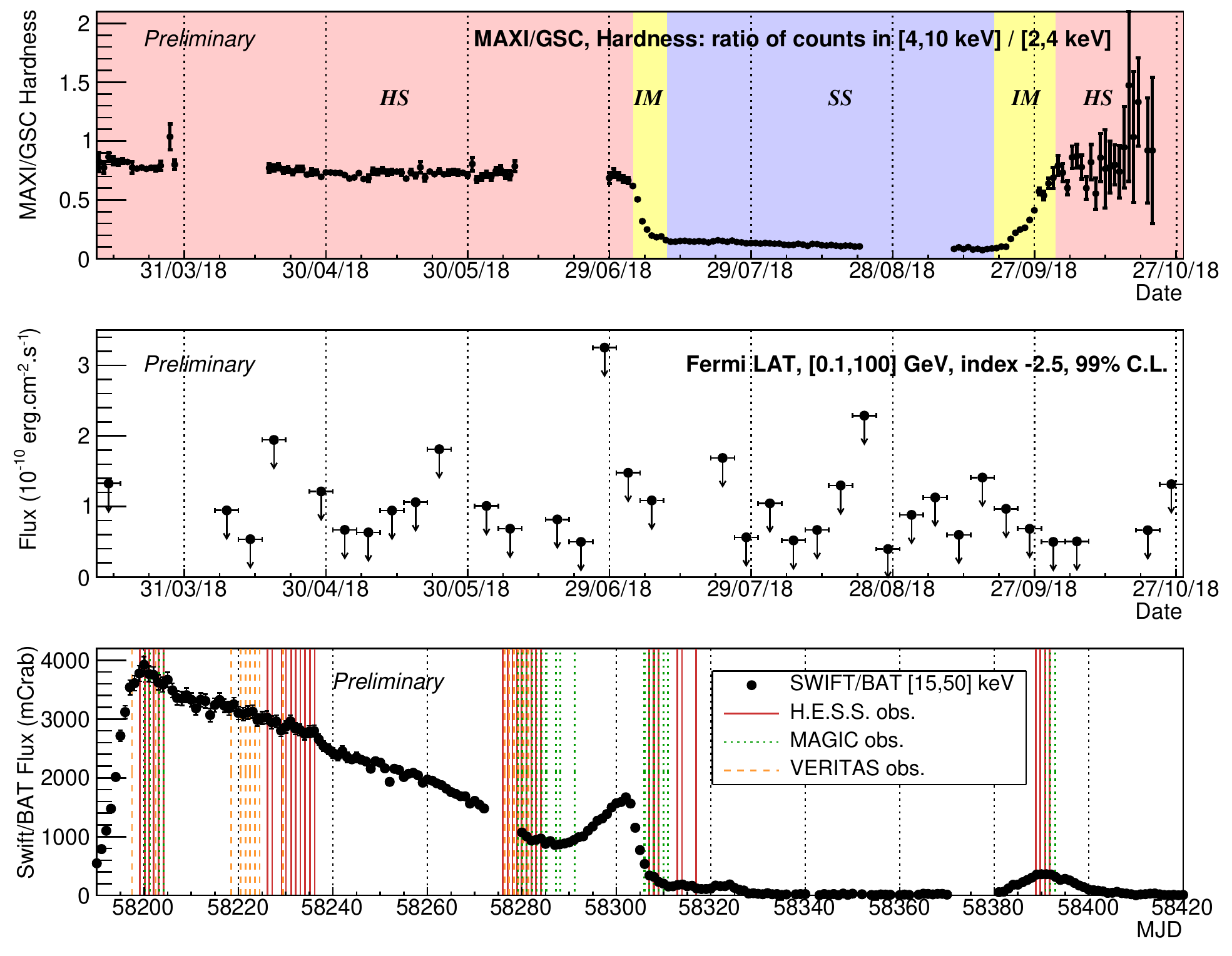}
    \caption{MAXI~J1820+070 data at different energy ranges. \textit{From top to bottom:} MAXI/GSC hardness ratio of the 4--10 to 2--4~keV fluxes\protect\footnotemark on which the states after~\cite{maxi_xray_optical} are superimposed (named here Hard State, Intermediate, and Soft State);  \textit{Fermi}-LAT\protect\footnotemark integral flux upper limits above 100~MeV with a 99\% confidence level and a power-law index of -2.5; \textit{Swift}/BAT\protect\footnotemark light curve in the 15--50~keV range  showing the nights in which the source was observed by H.E.S.S., MAGIC or VERITAS. For reference, MJD58200 corresponds to 23/03/2018.}
    \label{fig:mwl}
    \vspace*{-0.2cm}
\end{figure}

\begin{figure}
	\centering
    \includegraphics[width=0.78\linewidth]{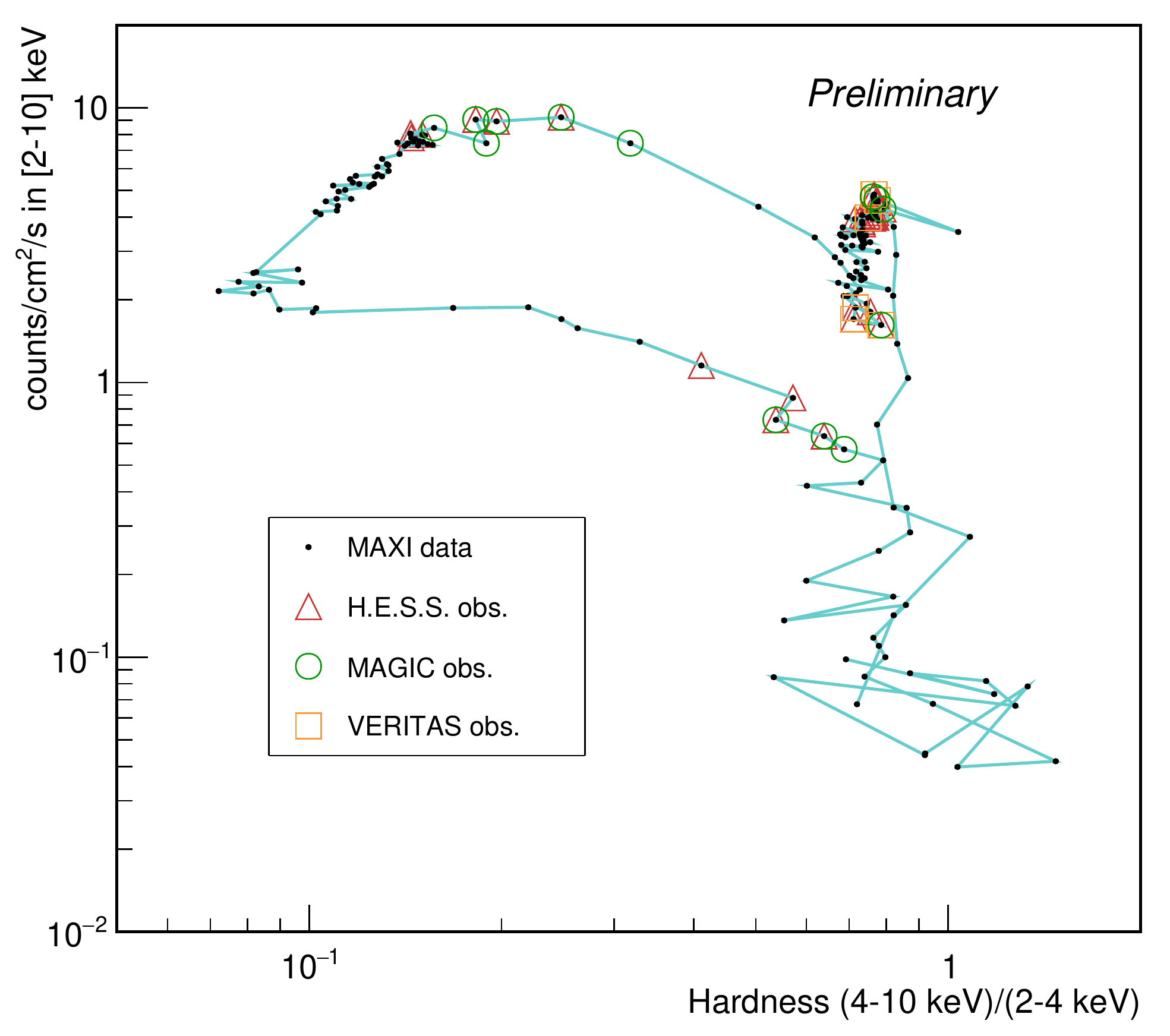}
    \caption{Hardness-intensity diagram obtained from MAXI/GSC X-ray observations in the 2--4 and 4--10 keV bands. The nights in which H.E.S.S., MAGIC or VERITAS observations were carried out have a triangle, a circle, and a square, respectively, around the MAXI point.}
    \label{fig:hid}
    \vspace*{-0.2cm}
\end{figure}

The bottom panel in Fig.~\ref{fig:mwl} shows the \textit{Swift}/BAT X-ray flux in the range 15--50~keV \cite{krimm13} from March to October 2018 to illustrate different states of MAXI~J1820+070 during its 2018 outburst period. The days in which H.E.S.S., MAGIC or VERITAS observations were conducted are shown as vertical lines. These days are also depicted together with the MAXI/GSC HID in Fig.~\ref{fig:hid}. The source was observed at VHE by H.E.S.S., MAGIC and VERITAS between March and June 2018, during the HS of the source. MAGIC and H.E.S.S. also observed it in July 2018, when the HS--SS transition took place~\cite{2018ATel11820....1H}, and in September-October 2018, during the SS--HS transition~\cite{2018ATel12064....1M}. The total exposure time after standard data quality selection cuts is 61.6~h (26.9~h H.E.S.S., 22.5~h MAGIC, 12.2~h VERITAS). All the observations were performed in wobble mode \cite{fomin94} and during dark time, with the Moon below the horizon. A summary of VHE observation conditions is shown in Table~\ref{tab:observation_conditions}. Optical observations with MAGIC's Central Pixel in M2 were also taken in ON mode (wobble offset = $0\degree$). Estimation of optical background was done using optical observations taken before and after MAXI~J1820+070 optical observation.  

%#######################################
\section{Data analysis and results}\label{results}
\vspace*{-0.1cm}

\subsection{VHE gamma rays}
\vspace*{-0.1cm}
The MAGIC data analysis followed the standard procedure described in \cite{magic16}.
%The shower images formed in the camera are parametrized through the Hillas parameters \cite{hillas85}, obtained after calibrating and cleaning the image from the noise generated by the night sky background and the electronics. This cleaning is done by setting constraints on the size and arrival time of the Cherenkov photons.
The discrimination between showers triggered by gamma-rays and hadrons is done through a random forest method \cite{magic08}, which is also used to reconstruct the gamma-ray arrival direction.
%The photon energy is estimated by comparing the Hillas parameters with look-up tables for a given shower intensity and distance from the shower axis to the telescopes (the so-called impact parameters).
The angular distance $\theta$ between the source position and the reconstructed gamma-ray direction is used to separate signal from background events, as the former are concentrated at small $\theta$ whereas the latter do not show any dependence on the angular distance. The H.E.S.S. analysis relies on a likelihood reconstruction technique based on a semi-analytical model \cite{2009APh....32..231D} providing gamma-rays/hadrons discriminating variables. It combines different configurations of telescopes \cite{2015arXiv150902896H}. Like in MAGIC analysis, the $\theta$ angular distance is used to cope with the background remaining after gamma-rays/hadrons discrimination. The VERITAS data analysis was performed following the standard analysis procedure~\cite{2008ICRC....3.1325D}. The analysis used box cuts optimized for a source with a soft spectrum~\cite{veritas_performance}. 

%% Be careful, this block might have to be moved elsewhere to get the footnotes on the same page as the figure of MWL data, if the figure moves
\addtocounter{footnote}{-3} %% to start at 4, like in the caption
\stepcounter{footnote}\footnotetext{Public data downloaded from \url{http://maxi.riken.jp/pubdata/v6l/J1820+071}}
\stepcounter{footnote}\footnotetext{Public data at \url{https://fermi.gsfc.nasa.gov/ssc/data/access/lat}}
\stepcounter{footnote}\footnotetext{Public data at \url{https://swift.gsfc.nasa.gov/results/transients/weak/MAXIJ1820p070}  }
%%%%%%%%%%%%%%%%%%%%%%%%%

The separate analyses performed by MAGIC, H.E.S.S and VERITAS yield no significant VHE emission from MAXI~J1820+070. The integral flux upper limits (ULs) for the whole data set at the 99\% confidence level obtained by each of the collaborations are shown in Table~\ref{tab:ul}. In order to compute these ULs, the gamma-ray spectrum has been assumed to follow a power-law with spectral index -2.5, and a common energy threshold of 182~GeV has been considered. No significant signal was detected at lower gamma-ray energies by \textit{Fermi}-LAT. The middle panel of Fig.~\ref{fig:mwl} shows the \textit{Fermi}-LAT ULs above 100~MeV with a 99\% confidence level and an assumed power-law index of -2.5, computed on 5-days time bins.

\begin{table}
 \begin{center}
	\caption{H.E.S.S., MAGIC and VERITAS integral flux upper limits for E > 182~GeV with an assumed spectral index of -2.5 and a confidence level of 99\%. The effective observation time for each collaboration is also shown.}
	\begin{tabular}{c c c}
    \hline \hline
	Instrument	& Effective time [h]	& UL [cm$^{-2}$ s$^{-1}$]   \\
    \hline
	H.E.S.S.	& 26.9		            & $1.1\times10^{-12}$       \\
    MAGIC       & 22.5 	                & $4.7\times10^{-12}$       \\
    VERITAS     & 12.2                  & $2.5\times10^{-12}$       \\
    \hline
    \end{tabular}
    \label{tab:ul}
 \end{center}
 \vspace*{-0.3cm}
\end{table}

\subsection{Optical}

MAGIC installed an alternative readout system in the central PMT of the MAGIC II camera, the Central Pixel, to perform simultaneous optical observations~\cite{cpix_icrc}. With this upgrade, MAGIC can observe short (1 ms--1 s) optical pulses. We searched for non-periodic optical pulses from MAXI J1820+070 that are brighter than 13 mag and last for less than 100 ms. To convert the Central Pixel output voltage to the corresponding optical absolute magnitude, we used the empirical expression mentioned in~\cite{cpix_icrc}. In order to improve sensitivity, an averaging filter of integration length of 1 ms was applied to de-noise the signal from the Central Pixel data (taken at the standard sampling rate of 10 kHz). At each point, the uncertainty is estimated as the standard deviation of each 1 ms window. The main source of irreducible optical background are faint meteors passing through the field of view of the central pixel, producing signals lasting about 5 to 20 ms. This background rate is low (frequency between 10$^{-3} \text{--} 10^{-5}$ Hz, decreasing with brightness). 
 
 A summary of MAGIC's optical observation and results is shown in Table~\ref{tab:MAGIC_Optical}. For the HS period, 2 epochs of optical observation were carried out, approximately 3 months apart. The optical activity remained consistently high in both epochs (23 count/1.5 h in March and 59 count/3 h in June). Shortly after, during the transition period that took place in early July, the activity dropped significantly (5 count/1 h), near the background level. No optical activity was detected when the source re-entered the HS state in September. The absence of optical events could be either due to the intrinsic source activity, or due to low statistics. For details of high optical activity during the HS, see \cite{maxi_xray_optical}.

 \begin{table}
 \begin{center}
	\caption{MAGIC optical observation surviving quality cuts and the corresponding results during different states of the source.}
	\begin{tabular}{c c c}
    \hline \hline
	Epoch	& Effective time [h]	& Optical activity [count/h]   \\
    \hline
	HS              & 4.5		    & 18.2    \\
    IM (HS-SS)      & 1             & 5       \\
    IM (SS-HS)      & 0.3           & 0       \\
    \hline
    \end{tabular}
    \label{tab:MAGIC_Optical}
 \end{center}
 \vspace*{-0.3cm}
\end{table}

%#######################################
\section{Discussion}\label{discussion}
\vspace*{-0.20cm}
MAXI~J1820+070 drew a significant multi-wavelength attention due to its exceptionally high X-ray flux. The detection of enhanced optical activity in the HS by the Central Pixel installed in MAGIC suggests that the optical flux could originate from jet emission, which is likely to be suppressed in the IM and SS. However, the source is not detected in gamma-rays above 100~MeV, highlighting the unclear relation between X- and gamma-ray emission \cite{fender01}. X-rays in microquasars mainly come from thermal disk emission and IC scattering with the corona, while gamma-rays are likely originated via IC scattering with the jet \cite{bosch09a}. Due to the different physical origins of the X- and gamma-ray luminosities, there are many environmental parameters affecting the relation between them. Some of the most influential ones are the magnetic field intensity, which regulates synchrotron losses in the jet, the orbital separation of the microquasar, and the fraction of jet energy dedicated to accelerate non-thermal particles. The upper limits at VHE presented in this work will provide valuable data to constrain some of these parameters and limit the possible physical scenarios.
%Two well known microquasars, Cygnus~X-1 and Cygnus~X-3, have been detected in gamma-rays \cite{zanin16,zdziarski18} showing a significantly lower X-ray emission than MAXI J1820+070 at its peak (a factor $\sim 3$ and $\sim 10$ in the \textit{Swift}/BAT range, respectively). Gamma-ray emission has likely a jet origin through inverse Compton (IC) scattering of stellar/disk/corona photons by relativistic electrons, although hadronic processes cannot be discarded \cite{bosch09a}. Being high-mass microquasars with a companion star above $\sim 10 M_\odot$, Cygnus~X-1 and Cygnus~X-3 IC emission is enhanced owing to the intense photon field produced by the star. This is not the case for MAXI~J1820+070 or any other LMMQ, in which the low mass companion may make IC radiation to only account for a small fraction of the total radiative output of the jets.

%#######################################
\vspace*{-0.20cm}
\acknowledgments
\vspace*{-0.20cm}
{\footnotesize
\textbf{MAGIC Collaboration:} We would like to thank the Instituto de Astrof\'isica de Canarias for the excellent working conditions at the Observatorio del Roque de los Muchachos in La Palma. The financial support of the German BMBF and MPG, the Italian INFN and INAF, the Swiss National Fund SNF, the ERDF under the Spanish MINECO (FPA2017-87859-P, FPA2017-85668-P, FPA2017-82729-C6-2-R, FPA2017-82729-C6-6-R, FPA2017-82729-C6-5-R, FPA2017-90566-REDC, AYA2015-71042-P, AYA2016-76012-C3-1-P, ESP2017-87055-C2-2-P), the Indian Department of Atomic Energy, the Japanese JSPS and MEXT, the Bulgarian Ministry of Education and Science, National RI Roadmap Project DO1-153/28.08.2018 and the Academy of Finland grant nr. 320045 is gratefully acknowledged. This work was also supported by the Spanish Centro de Excelencia "Severo Ochoa" SEV-2016-0588 and SEV-2015-0548, and Unidad de Excelencia "Mar\'ia de Maeztu" MDM-2014-0369, by the Croatian Science Foundation (HrZZ) Project IP-2016-06-9782 and the University of Rijeka Project 13.12.1.3.02, by the DFG Collaborative Research Centers SFB823/C4 and SFB876/C3, the Polish National Research Centre grant UMO-2016/22/M/ST9/00382 and by the Brazilian MCTIC, CNPq and FAPERJ.

\textbf{H.E.S.S. Collaboration:} The support of the Namibian authorities and of the University of Namibia in facilitating the construction and operation of H.E.S.S. is gratefully acknowledged, as is the support by the German Ministry for Education and Research (BMBF), the Max Planck Society, the German Research Foundation (DFG), the Helmholtz Association, the Alexander von Humboldt Foundation, the French Ministry of Higher Education, Research and Innovation, the Centre National de la Recherche Scientifique (CNRS/IN2P3 and CNRS/INSU), the Commissariat \`a l'\'energie atomique et aux \'energies alternatives (CEA), the U.K. Science and Technology Facilities Council (STFC), the Knut and Alice Wallenberg Foundation, the National Science Centre, Poland grant no. 2016/22/M/ST9/00382, the South African Department of Science and Technology and National Research Foundation, the University of Namibia, the National Commission on Research, Science \& Technology of Namibia (NCRST), the Austrian Federal Ministry of Education, Science and Research and the Austrian Science Fund (FWF), the Australian Research Council (ARC), the Japan Society for the Promotion of Science and by the University of Amsterdam. We appreciate the excellent work of the technical support staff in Berlin, Zeuthen, Heidelberg, Palaiseau, Paris, Saclay, T\"ubingen and in Namibia in the construction and operation of the equipment. This work benefitted from services provided by the H.E.S.S. Virtual Organisation, supported by the national resource providers of the EGI Federation.

\textbf{VERITAS Collaboration:} This research is supported by grants from the U.S. Department of Energy Office of Science, the U.S. National Science Foundation and the Smithsonian Institution, and by NSERC in Canada. This research used resources provided by the Open Science Grid, which is supported by the National Science Foundation and the U.S. Department of Energy's Office of Science, and resources of the National Energy Research Scientific Computing Center (NERSC), a U.S. Department of Energy Office of Science User Facility operated under Contract No. DE-AC02-05CH11231. We acknowledge the excellent work of the technical support staff at the Fred Lawrence Whipple Observatory and at the collaborating institutions in the construction and operation of the instrument.

\vspace*{0.20cm}

We would like to thank Sera Markoff, Phil Uttley, and our colleagues from the X-ray, Optical/Infrared, and Radio communities, for the very fruitful exchanges we had on MAXI~J1820+070.
}

\vspace*{-0.20cm}
\FloatBarrier
\bibliographystyle{JHEP}
\bibliography{refs}

\end{document}